\documentclass[5p,times]{elsarticle}

\usepackage{amssymb}
\usepackage{booktabs}
\usepackage{siunitx}
\usepackage{lscape}
\usepackage{multirow}
\usepackage{tikz}
\usepackage{colortbl}
\usepackage{array}
\usepackage{svg}
\usepackage{graphicx} 
\usepackage{adjustbox}
\usepackage[breaklinks=true]{hyperref}

\newcolumntype{L}[1]{>{\raggedright\let\newline\\\arraybackslash\hspace{0pt}}m{#1}}
\newcolumntype{C}[1]{>{\centering\let\newline\\\arraybackslash\hspace{0pt}}m{#1}}
\newcolumntype{R}[1]{>{\raggedleft\let\newline\\\arraybackslash\hspace{0pt}}m{#1}}

\newcommand\blfootnote[1]{%
  \begingroup
  \renewcommand\thefootnote{}\footnote{#1}%
  \addtocounter{footnote}{-1}%
  \endgroup
}

\usepackage{soul}
\usepackage{color}
	\definecolor{celadon}{rgb}{0.67, 0.88, 0.69}
    \definecolor{flamingopink}{rgb}{0.99, 0.56, 0.67}

\usepackage{pbox}
\usepackage{makecell}
\usepackage{appendix}

\usepackage{footnote}
\usepackage{tablefootnote}

\usepackage{graphicx}
\usepackage{tabularray}
\usepackage{xcolor}

\definecolor{gray7}{HTML}{BFBFBF}
\definecolor{gray9}{HTML}{E5E5E5}

\usepackage{tabularray}

\makesavenoteenv{tabular}
\makesavenoteenv{table}

\journal{Forensic Science International: Digital Investigation}

\begin{document}
\emergencystretch 3em

\begin{frontmatter}



\title{Exploring the Potential of Large Language Models for Improving Digital Forensic Investigation Efficiency}

\renewcommand{\theaffn}{\arabic{affn}}

\author[label1]{Akila Wickramasekara}
\author[label2,label3]{Frank Breitinger}
\author[label1]{Mark Scanlon}

\affiliation[label1]{organization={Forensics and Security Research Group, School of Computer Science},
            addressline={University College Dublin},
            city={Belfield},
            state={Dublin 4},
            country={Ireland}}

\affiliation[label2]{organization={School of Criminal Justice},
             addressline={University of Lausanne},
             state={Lausanne},
             country={Switzerland}}

\affiliation[label3]{organization={Present affiliation: Institute of Computer Science},
             addressline={University of Augsburg},
             state={Augsburg},
             country={Germany}}




    

\begin{abstract}
  The ever-increasing workload of digital forensic labs raises concerns about law enforcement's ability to conduct both cyber-related and non-cyber-related investigations promptly. Consequently, this article explores the potential and usefulness of integrating Large Language Models (LLMs) into digital forensic investigations to address challenges such as bias, explainability, censorship, resource-intensive infrastructure, and ethical and legal considerations. ~A comprehensive literature review is carried out, encompassing existing digital forensic models, tools, LLMs, deep learning techniques, and the use of LLMs in investigations. 
The review identifies current challenges within existing digital forensic processes and explores both the obstacles and the possibilities of incorporating LLMs. In conclusion, the study states that the adoption of LLMs in digital forensics, with appropriate constraints, has the potential to improve investigation efficiency, improve traceability, and alleviate the technical and judicial barriers faced by law enforcement entities.
\end{abstract}




\end{frontmatter}


\section{Introduction}

With the widespread growth of information and communication technology (ICT) and information systems, cybercrimes have seen a significant increase in recent years~\cite{ali2019cyber}\footnote{https://go.crowdstrike.com/rs/281-OBQ-266/images/GlobalThreatReport2024.pdf}. As a further compounding factor, the number of ``traditional'' police investigations that include digital evidence is also increasing~\cite{10.1145/3339252.3340517}. Addressing and investigating this volume of cases presents substantial challenges. 

Generative AI (GenAI) and Large Language Models (LLMs) have become prominent topics of global discussion, prompting researchers to intensify their investigations by leveraging the capabilities of LLMs. The usage of LLMs within the scientific community experienced a rapid surge after 2022, notably with the advent of OpenAI's ChatGPT platform. In a relatively short period of time, this topic has attracted the attention of academia, industry, and the research community at large~\cite{Partha2023ChatGPT}. Simultaneously, researchers are exploring the potential of LLMs in various domains and assessing their impact on the future of science and society. This inquiry also includes an examination of the potential harmfulness associated with the deployment of LLMs~\cite{Larochelle2020LanguageFewShot,nozza2022measuring}. In other words, the use of LLMs in various tasks can be a double-edged sword, necessitating careful consideration depending on the specific situations and contexts.

Given the rapidly evolving landscape of LLMs, it is prudent to look into various types and their unique capabilities. A nuanced understanding of the strengths and characteristics of different LLMs can contribute to more informed and effective applications within the dynamic field of digital forensics (DF).

This paper reviews recent advances in the application of LLMs within digital forensics, focussing on established models, methods, and key challenges. By examining contemporary studies from 2019 onwards, the survey highlights core areas, such as automation, investigative methods, and efficiency improvements facilitated by LLMs. In addition, it explores the literature that addresses challenges in both DF and LLMs, covering limitations, ethical considerations, and forensic-specific risks. This comprehensive review synthesises current insights and emerging trends, offering a foundation for understanding the potential and limitations of LLMs in DF contexts.

In light of fast-paced advancements and the recent explosion in LLM-focused research, a substantial influx of LLM-focused research papers has occurred since the launch of ChatGPT in late 2022. Due to this fast pace, many seminal research articles exist solely as preprints on preprint services, e.g., arXiv. To give two examples, the initial papers for GPT-4~\cite{openai2023gpt4} and LLaMA~\cite{touvron2023llama} are only published on arXiv, but have garnered thousands of citations each. The recent release of DeepSeek-V3 in late 2024 and the corresponding technical reqport~\citep{deepseekai2024} is also only available on arXiv. Despite their preprint status, these articles offer essential insights critical for contemporary research and dialogue within the domain, making their incorporation into this article necessary to provide the most up-to-date knowledge and perspectives.

The paper is structured as follows: Section~\ref{background} provides a comprehensive background for the review, delving into existing DF process models, the challenges inherent in DF, and a detailed overview of the current work conducted with the use of LLMs within DF. In Section~\ref{LLMS} the paper delves into the realm of Natural Language Processing (NLP), elucidating the working principles of LLMs, their architectural foundations, and the specifics surrounding specially trained LLMs. Section~\ref{LLMS_capabilities} provides an in-depth review that focusses on the capabilities and benchmark information of LLMs trained for coding tasks, as well as those tailored for vision assistance. Section~\ref{LLMS_capabilities} explores the synergy between DF and LLMs, detailing how LLMs can be effectively employed in each phase of the DF process model. Finally, in Section~\ref{conclution}, the paper summarises the future challenges associated with integrating LLMs with automated agents within the DF domain. The conclusion outlines potential avenues for future research and development, shedding light on the path of future DF investigations employing LLMs. The discussion covers not only the potential negative impacts but also the practical difficulties and risks in real world environments.

\subsection{Digital Forensic Context}
\label{background}

DF is a process for identifying, preserving, analysing, and documenting digitally recorded data, which originate in electronic devices such as computers, servers, smartphones, and IoT devices~\cite{baryamureeba2004enhanced}. This exercise is required in most criminal cases. Data collected in this process are kept unchanged and safe for presentation in a court case or to support future investigations conducted by law enforcement agencies~\cite{mukherjee2018review}.

\subsubsection{Digital Forensic Process Models}
\label{DF_process_models}

DF process models consist of a series of activities that help standardise the investigative process~\cite{du2017evaluation} and outline the phases; collection, preservation of evidence, examination or analysis, and reporting. 

\begin{figure}
    \centering
    \includegraphics[width=\columnwidth]{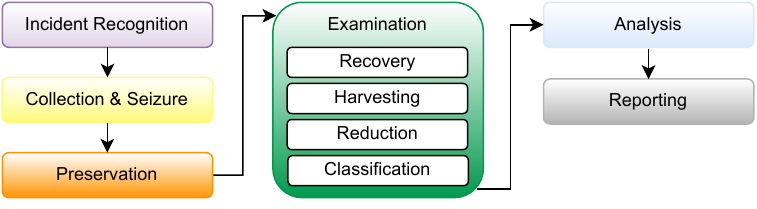}
    \caption{Traditional digital forensic process model~\cite{casey2011digital}}
    \label{fig:general-df-model}
\end{figure}

DF encompasses various subdisciplines such as computer forensics, mobile device forensics, memory forensics, network forensics, and cloud forensics, each employing distinct processes reflected in a plethora of models within the literature~\cite{2023SCANLONChatGPT,wu2020digital}. These models often share phases but differ in their focus and execution. For example, \citet{al2020review} proposed a mobile forensic model that adds a preparatory phase and bifurcates the analysis stage into examination and analysis phases. To accommodate the complexities of computer, network, cloud, and smart device forensics, \citet{lutui2016multidisciplinary} introduced a multidisciplinary model that requires diverse skills for effective investigation, ranging from incident detection to evidence storage.

Casey's model, as shown in Figure~\ref{fig:general-df-model}, includes phases such as incident recognition, evidence collection, preservation, and presentation, with the examination phase detailed in recovery, harvesting, reduction, and classification~\cite{casey2011digital}. During incident recognition, the focus is on identifying the incident itself, possible evidentiary sources, and expected digital evidence types, as well as delineating the scope of the ensuing investigation. Conversely, investigators systematically acquire pertinent evidence from various sources encompassing computers, smartphones, storage media, and networks during collection and seizure. Preservation is of paramount importance in upholding the integrity of evidence, necessitating specific and accurate measures to ensure the unmodified condition of the collected data throughout the investigative process. The overarching objective remains the meticulous safeguarding of evidence integrity.

The subsequent phase entails examination, in which analysts rigorously scrutinise the gathered data to extract pertinent information. This endeavour may involve the use of various forensic hardware and software tools and techniques. The examination process includes the interpretation of the information extracted to draw conclusive inferences about the events under scrutiny. This phase often demands a profound understanding of both the technology employed and the context surrounding the evidence.

Next is the reporting phase, where the findings derived from the analysis are presented systematically in a format suitable for legal adjudication. This may involve preparing comprehensive reports and providing expert testimony in a court of law. This model emphasises the critical nature of maintaining the integrity and provenance of the evidence and the requirement for expert analysis to extract and interpret pertinent information. This culminates in a report suitable for legal scrutiny. In particular, the analysis or examination stage is crucial in all models, demanding specialised knowledge in the relevant DF area~\cite{mir2016analysis,prayudi2020framework}.

The advent of cloud computing has led to the Digital Forensic as a Service (DFaaS) model by \citeauthor{van2014digital}, which integrates evidence preservation and analysis into an automated and secure software service, marking a significant evolution in forensic methodologies~\cite{2015VANBEEKDFaaS,du2017evaluation,van2020digital}.

\subsubsection{Existing Challenges in Digital Forensics}

DF is an evolving field, yet the literature highlights that it still undergoes changes to address ongoing challenges and advancements. \citet{dubey2023digital} assert that DF faces key challenges, including the complexity of data and its volume, a lack of standardisation, inadequacies in the power of existing tools to support investigations, and issues related to timelines.


In addition to the previously mentioned challenges, other issues persist including scope creep in cases due to complexity and the vast data involved, selecting and prioritising the right set of evidence, and efficiently allocating time and investigators for the chosen evidence~\cite{Kalaimannan2013Maximizing}. 
\citet{Koper2014Optimizing} focus on a number of these issues from the investigator's perspective, including challenges in adapting to a system, unexpected time sinks, and frustrations among officers arising from expected operational timelines and the adoption of complex systems. The contemporary landscape of forensic science is characterised by a shortage of proficient agents, exacerbated by the swiftly evolving standards, practices, tools, and techniques within the field. Moreover, the predominant emphasis of law enforcement roles on fieldwork, as opposed to dedicated DF duties, has further contributed to the scarcity of adept human expertise in this domain~\cite{Eva2016Challenges}.

Automating the DF process using existing technology appears to be a promising solution to address issues related to time management and effectiveness~\cite{michelet2023automation}. However, an ongoing challenge revolves around measuring the accuracy of investigations and ensuring the verification of the automated process. This aspect remains an open area that requires more attention and resolution~\cite{Jarrett2021Impact}.

\subsubsection{Existing Work With LLMs in Digital Forensics}
\label{exsisting_workDF}
\citet{2023SCANLONChatGPT} analysed using ChatGPT for DF. In their assessment, the authors evaluated the programming, incident narration, keyword list creation, and DF teaching abilities of ChatGPT. Their conclusion highlighted that while ChatGPT exhibited some hallucinations in the output results, it still serves as an effective assistant for code generation. \citet{wickramasekara2025} introduced the AutoDFBench benchmarking framework, and corresponding score, to evaluate code generation for DF specific tasks against the tests and datasets used as part of NIST's Computer Forensics Tool Testing Program (CFTT)\footnote{\url{https://www.nist.gov/itl/ssd/software-quality-group/computer-forensics-tool-testing-program-cftt}}.

Timeline reconstruction helps investigators deduce the chronological ``story'' of an event. In line with timeline regeneration, \citet{2023SilalahiLLMDrone} proposed a method to detect anomalies in a drone flight by employing sentiment analysis with the assistance of a pre-trained LLM. Their approach successfully discerned the differences between normal and abnormal events with an accuracy of 92.5\%. 

Hansken, a DFaaS platform created by the Netherlands Forensic Institute, is designed to help investigators handle evidence and conduct investigations more efficiently~\citep{2015VANBEEKDFaaS}. ChatGPT has been used as an assistance for the Hansken DFaaS system using its bespoke query language, contributing to streamlined processes and improved support for investigators. In these experiments by \citet{henseler2023chatgpt}, it was tasked
with analysing evidence using Hansken's trace model. This work demonstrated the potential for ChatGPT in helping with analytical aspects of investigations, highlighting its ability to process and interpret evidence data.

While DF is the main focus of this paper, it is important to recognise the broader application of LLMs in adjacent areas within cybersecurity, many of which overlap with DF. LLMs are proving to be valuable tools in fields such as malware analysis, security log analysis, code security reviews, and intrusion detection areas that bridge the gap between cybersecurity and DF~\cite{Yu2024maltrack, karlsen2024benchmarking, li2023myriad}. In malware analysis, LLMs can identify patterns in malicious code, while in log analysis, they assist in detecting anomalies across large datasets, thereby improving response times. In code-related security reviews, LLMs like GRACE have demonstrated the ability to identify vulnerabilities in software, achieving a detection rate of 28.65\% of vulnerabilities~\cite{LU2024112031}. These applications contribute to DF investigations by improving and improving evidence collection and analysis.

\section{Background of Large Language Models}
\label{LLMS}

This section explores LLMs, concentrating on three principal aspects. Initially, it explores the architecture of LLMs, detailing their design and function. Then it assesses the usability of LLMs, underscoring the features and capabilities that make them suitable for a wide range of tasks. Finally, it showcases the versatility of LLMs by discussing their applications across various fields, demonstrating their wide-reaching impact and the extensive scope of their applications.

\subsection{Natural Language Processing}

Popular LLMs such as Generative Pre-trained Transformer (GPT)~\citep{dale2021gpt}, Language Model for Dialogue Applications (LaMDA)~\citep{thoppilan2022lamda}, Pathways Language Model (PaLM)~\citep{chowdhery2022palm}, Bidirectional Encoder Representations from Transformers (BERT)~\citep{devlin2018bert}, and Large Language Model Meta AI (LLaMA)~\citep{touvron2023llama} stem from advances in Natural Language Processing (NLP). NLP, which focusses on language-based tasks, uses traditional and deep learning models to enable applications such as language translation, text processing, and speech recognition \cite{reshamwala2013review}.

Deep learning, a branch of machine learning, uses complex computational layers and adaptive weights to improve prediction accuracy, offering a more refined analysis than conventional machine learning~\cite{lecun2015deep}. It has excelled in image and speech recognition and natural language understanding, mimicking the decision-making process of the human brain through artificial neurons. These neurons form networks capable of intricate pattern recognition and data analysis. Central to deep learning are neural networks with multiple hidden layers that autonomously learn and extract features from data, bypassing the need for manual variable selection. This automatic feature extraction makes them exceptionally adept at handling complex tasks~\cite{DONG2021100379}.

\subsection{LLMs}
An LLM is a language model that employs neural networks with billions of parameters, trained on extensive text data. These models are engineered to comprehend and generate human language. Fundamentally, they rely on multiple neural network architectures, enabling them to recognise the relationships between words and phrases within sentences~\cite{Yang2024PowerofLLMs,shen2023chatgpt}. These architectures have been a transformative force in natural language processing. Its capability to excel across a diverse array of language-related tasks distinguishes it as a game-changer, in contrast to being tailored for a singular, specific task.

\subsection{Architecture of LLMs}


LLMs utilise deep learning, particularly neural networks, to process and produce human language. Fundamentally, a language model operates with letters or words, but since machine learning algorithms and neural networks require vector inputs, words are vectorised. Each word in the vocabulary is assigned a unique numerical value for input into neural networks. Through initial random weight assignments and subsequent backpropagation, words acquire numerical positions reflecting their semantic similarity, culminating in a word embedding model~\cite{lai2016generate}.

\subsubsection{Word to Vectors}
Word embeddings, as introduced by \citet{mikolov2013efficient}, entail precise and high-dimensional vector representations for words, particularly suited for extensive datasets comprising billions of text entries. The authors explored model architectures for word vectorisation, achieving substantial improvements in accuracy while requiring fewer computational resources and reduced training time~\cite{mikolov2013efficient}. In the realm of LLMs, the primary objective is to generate new text based on the extensive dataset on which it was trained. For this purpose, \citet{vaswani2017attention} introduced the transformer model, assisted by the word-to-vector model. This architecture incorporates a self-attention mechanism, as well as encoder and decoder processes, enabling the model to quickly and simultaneously focus on pertinent information.

\subsubsection{Transformer Models}

The transformer model initially aimed at machine translation, translating input words into another language, begins with word embedding, where input, termed tokens, are vectorised. Recognising word order is achieved through positional encoding, with two main techniques: absolute and relative. Absolute positional encoding assigns unique vectors to each position, enhancing the model's ability to recognise word placement and facilitate position-specific attention~\cite{ke2020rethinking}. Relative positional encoding, on the other hand, calculates the relative positions of words by introducing a bias term that quantifies the distances between positions, improving the model's ability to understand the relationships between words within a sequence~\cite{ke2020rethinking}.

Self-attention, a core mechanism within the transformer, calculates the relationship among words in a sentence, allowing the model to assess the similarity of each word with others and generate unique representations for each~\cite{al2019character}. The decoder mirrors the encoder's steps but uses different weights, starting with positional encoding and computing self-attention values to identify the sentence's initial translation word.

This transformer process, which leverages stacking self-attention and unique positional encoding, has significantly advanced NLP tasks, including machine translation, text generation, and summarisation, by executing these processes in parallel and optimising the weights of both the encoder and decoder~\cite{vaswani2017attention, acheampong2021transformer}.

\subsection{Specifically Trained LLMs}
\label{sec:Specifically_trained_LLMs}

The transformer model and the self-attention mechanism have paved the way for researchers to train language models on trillions of tokens with billions of parameters. Several LLMs have been trained and harnessed, each designed with specific capabilities for diverse fields such as security~\citep{li2023myriad}, chemistry~\citep{xuan2023llms,tsai2023ExploringChem}, engineering~\citep{hou2023large}, medicine~\citep{chang2023survey}, business~\citep{vidgof2023large}, tourism~\citep{carvalho2023chatgpt}, and language-related applications~\citep{chang2023survey}.
These models are used in tasks ranging from detecting security threats~\citep{li2023myriad}, analysing data and generating synthetic actions to teaching~\citep{moore2023empowering}, code generation~\citep{hou2023large,liu2023code}, structured query generation~\citep{vidgof2023large,Li2024text2sql}, planning~\citep{vidgof2023large}, assisting in medical education~\citep{chang2023survey}, clinical decision-making~\citep{eggmann2023implications}, leveraging clinical settings~\citep{thirunavukarasu2023large}, clinical validation~\citep{karabacak2023embracing}, understanding general patterns and decision-making~\citep{vidgof2023large}, bias detection~\citep{moore2023empowering}, addressing
ethical issues~\citep{bonner2023large}, language translations~\citep{caines2023application}, question answering~\citep{moore2023empowering}, information
extraction~\citep{Liu2024Information}, and business process automation~\citep{vidgof2023large}, among others~\citep{chang2023survey}. The fine-tuning and retraining capabilities of LLMs enable them to be adapted to specific tasks or behaviours in a predefined manner. Fine-tuning involves taking an already trained language model and retraining its existing weights and bias values using a new dataset specific to a particular domain. This process allows the LLM to be customised and refined for tasks beyond its original training, enhancing its applicability in specific contexts~\cite{ziegler2019fine}. This process results in a new model that is more tailored and focused on the specified domain. In the existing literature, it is frequently observed that LLMs are fine-tuned with a particular emphasis on engineering and research-related fields. This targeted fine-tuning ensures that the model is adept at handling tasks and generating content specifically relevant to the intricacies of these domains~\cite{chang2023survey}.

\subsection{Multi-modal Large Language Models (MLLMs)}

Unlike traditional LLMs, which are trained on text data, Multi-modal Large Language Models (MLLMs) are designed to process and interpret image-based data alongside text. These models can extract and analyse information within images or videos, integrating visual and textual data to enhance comprehension and analysis~\cite{Tan_2024_CVPR}. The application of MLLMs is rapidly expanding across fields such as digital forensics, where they can be used to analyse images of documents or identify visual anomalies. Section~\ref{vision} provides a discussion on the potential and diverse applications of MLLMs in various domains.

\subsection{Large Action Models}

  While LLMs excel in text generation and processing, they struggle with complex task manipulation and operational control, especially when moving from language understanding to action execution. This limitation arises because their core design emphasises prediction and generation over direct task execution. To overcome this shortcoming, recent research has introduced innovative approaches, such as the Large Action Model (LAM) developed by the Rabbit research team. LAM extends the capabilities of LLMs by incorporating action-based operations\footnote{https://www.rabbit.tech/research}. These LAMs can mimic human routines such as scheduling meetings with given instructions, sending emails, ordering taxis, and handling complex tasks such as making reservations for a whole trip.  In this approach, the base model is trained to comprehend sequences of human-provided actions and commands, allowing it to execute these actions and tasks accordingly. Similarly, Microsoft introduced the concept of Visualisation-of-Thought (VoT) aimed at integrating human cognitive abilities, specifically the creation of mental images, into the model~\cite{Wu2024MindsEO}. Through this approach, it has been demonstrated that MLLMs excel in visual tasks, thereby enabling the extension of action capabilities within an LAM to any LLM. These advancements signify promising directions toward enhancing the practical applicability and versatility of language models across various domains.

\section{Capabilities of Large Language Models}
\label{LLMS_capabilities}
 
This section focusses on the abilities and capabilities of Language Model Models (LLMs) as outlined in Section~\ref{sec:Specifically_trained_LLMs}. This section also discusses the currently available fine-tuned LLMs that exhibit potential for application in DF.   Although considered too broad for this article, \citet{10.1145/3639372} provide a detailed generic overview of LLMs, their operation, and how they are trained and fine-tuned.

\subsection{Programming/Code Generation}

The ability to generate source code within a specific context is a crucial skill inherent in a language model~\cite{alon2020structural}. \citet{Xu2022evaluationLLM} conducted a systematic evaluation of six LLMs for code generation in 12 different programming languages.   The benchmarking process employed the HumanEval benchmark, along with a tailored evaluation dataset designed to assess the functional correctness of the programs generated by an LLM~\cite{chen2021evaluating}. The benchmark comprises a set of coding problems in which the model is tasked with generating Python functions. Each problem is accompanied by a prompt and a set of unit tests that verify whether the generated code produces the expected output. This facilitates a systematic evaluation to generate both syntactically correct and functionally accurate programs. Using this dataset, it is possible to measure performance on real-world coding tasks, as well as its ability to generate solutions that satisfy functional requirements. 

 The Mostly Basic Programming Problems (MBPP) is another benchmark comprising 974 programming tasks. It serves as a frequently used evaluation dataset for LLMs specialising in code-related tasks~\cite{wei2022chain}.
Several LLMs explicitly trained for code generation include Code LLaMA~\citep{roziere2023code}, CodeGen~\citep{nijkamp2022codegen}, StarCoder~\citep{li2023starcoder}, PanGu-Coder~\citep{christopoulou2022pangu}, PanGu-Coder2~\citep{christopoulou2022pangu}, WizardCoder~\citep{luo2023wizardcoder}, InCoder 6B~\citep{fried2022incoder}, CodeGen-Mono 16B~\citep{nijkamp2022codegen}, Code-Davinci-001~\citep{zhou2022least},  Code-Davinci-002~\citep{zhou2022least}, PaLM-Coder-540B~\citep{chowdhery2022palm}, CodeT5+~\citep{wang2023codet5,wang-etal-2021-codet5}, InstructCodeT5+~\citep{chen2023codet,wang-etal-2021-codet5}, GPT-4 with Reflexion~\citep{shinn2023reflexion}, CodeGeeX~\citep{zheng2023codegeex}, AlphaCode~\citep{li2022competition}, Santa-Coder~\citep{Loubna2022Sanata}, CodeFuse-13B~\citep{Di2024CodeFuse}, Codex~\citep{KALYAN2024100048}, WaveCoder~\citep{yu-etal-2024-wavecoder}. A higher value for both HumanEval and MBPP indicates greater precision in code generation for a given task. For detailed information, refer to Table~\ref{summery_code_LLM}, which presents the counts for HumanEval and MBPP, along with the trained parameter size for each LLM. A higher score for both HumanEval and MBPP indicates greater precision
in code generation for a given task. 

Table~\ref{summery_code_LLM} presents the scores for HumanEval and MBPP for each of the code generation LLMs mentioned above, along with the trained parameter size for each LLM. Four generic LLMs, or Mixture-of-Experts (MoE) models, are also included in Table~\ref{summery_code_LLM}: o1-mini~\citep{yu2024humanevalprombpppro}, GPT-4 with Reflexion, DeepSeek-V3 and DeepSeek-V3-Base~\citep{deepseekai2024}. These are included as these are the top 4 best performing models for HumanEval despite them being MoE models.\blfootnote{* Estimated parameter count as value is not officially released~\cite{RIZZO202470}.}
\begin{table}
\centering
\caption{Trained parameter count, HumanEval and MBPP scores for LLM based code generation (ordered by HumanEval score).}
\label{summery_code_LLM}
\resizebox{.47\textwidth}{!}{
    \begin{tblr}{
      hlines = {0.05em},
      hline{1} = {-}{0.16em},
      hline{2} = {-}{0.16em},
      hline{3,20} = {-}{0.08em},
    }
    \textbf{Model}                    & \textbf{Parameters} & \textbf{HumanEval} & \textbf{MBPP}  \\
    o1-mini & 100B  &  97.6  &  93.9
    \\
    GPT-4 with Reflexion   & 1.76T*  &  91.0  &  77.1
\\
DeepSeek-V3 &    671B  &  85.6 &  -
\\
DeepSeek-V3-Base &    671B  &  65.2 &  75.4
\\
Code LLaMA &   34B&    62.2 &   61.2
\\
PanGu-Coder2&    15B  &  61.64 &  -
\\
WizardCoder &   15B  &  57.3  &  51.8
\\
Code-Davinci-002 (GPT3.5) &   175B &   47.0  &  58.1
\\
StarCoder &   15.5B   & 40.8  &  49.5
\\
Code-Davinci-001 (GPT3)  &  175B  &  39.0   & 51.8
\\
PaLM-Coder &   540B  &  36.0  & 47.0
\\
InstructCodeT5+   & 16B &   35.0 &   -
\\
code-cushman-001 &   12B  &  33.5   & 45.9
\\
CodeT5+  &  16B  &  30.9 &  -
\\
CodeGen-MONO &   16.1B  &  29.7 &   42.4
\\
CodeGen &   16.1B &   29.28  &  35.28
\\
Codex-12B  &  12B  &  28.81  &     -
\\
PanGu-Coder&    2.6B &   27.78 &   23
\\
Sanata-Coder  &  1.1B  &  18  &  35
\\
AlphaCode &   1.1B   & 17.1 &   -
\\
InCoder 6B  &  6.7B  &  16.4  &  21.3
    \end{tblr}%
}
\end{table}

\subsection{Vision Assistance}
\label{vision}

Traditional vision assistant systems face limitations in image processing or recognition, as they are typically trained on fixed types of datasets. However, with the emergence of LLMs, this paradigm has changed to the use of raw text as a source of supervision~\cite{Radford2021LearningVisual,Tewel_2022_CVPR}. Research on visual recognition language models is experiencing exponential growth, with the number of models exceeding 1,500 in 2023~\cite{zhang2023visionlanguage}. \citet{Radford2021LearningVisual} introduced a novel method called Contrastive Language Image Pre-training (CLIP). This method is efficient and capable of performing a wide range of tasks during pre-training. It enables a model to learn a shared representation space for both images and text, facilitating a deeper understanding of the relationships between the two modalities. \citet{pmlr-v139-ramesh21a} proposes a model for text-to-image generation, capable of generating images as combinations derived from textual input or sentences. Moreover, with the model named Generating Images with Large Language Models (GILL), it becomes feasible to generate text, retrieve images, generate novel images, and interleave the results into coherent multimodal dialogues~\cite{koh2023generating}.
VisionLLM is a framework leveraging LLMs for diverse vision tasks with unified language instruction, demonstrating generality and flexibility~\cite{wang2023visionllm}. It incorporates a language-guided image tokeniser and an LLM-based task decoder, capable of handling open-ended tasks based on provided language instructions~\cite{wang2023visionllm}.

Visual instruction tuning leverages language-only models, such as GPT-4, to generate multimodal language-image instruction following data. This data is then utilised to instruction-tune large multimodal models, such as Large Language and Vision Assistant (LLaVA)~\cite{liu2023visual,liu2023improved, openai2023gpt4}. The open source LLaVA project introduces an end-to-end trained model, integrating a vision encoder with an LLM. Notably, LLaVA showcases multimodal chat capabilities. LLaVa has the capability to interact with images, provide detailed descriptions and respond to queries with a reported accuracy of 92.53\%~\cite{liu2023visual}. This shows its effectiveness in understanding and generating contextually relevant information on visual content~\cite{liu2023visual}. 
MiniGPT-4 is an open-source, powerful visual instruction-tuned LLM, and it demonstrates versatility by generating stories and poems inspired by provided images and teaching users how to cook based on visual cues from food photos. This showcases its ability to understand and respond creatively to various visual stimuli~\cite{zhu2023minigpt4}.

Position-Enhanced Visual Instruction Tuning (PVIT) represents an extended version of Multimodal Large Language Models (MLLMs). It facilitates region-level encoding in an image, enabling the model to discern and identify information within specific regions~\cite{chen2023positionenhanced}.  This model enables users to interact with both the language and drawing the bounding boxes to indicate the area of interest within an image~\cite{chen2023positionenhanced}.
Other MLLMs, such as Visual ChatGPT~\citep{Radford2021LearningVisual}, InternGPT~\citep{liu2023interngpt}, Flamingo~\citep{Alayrac2022Advances}, BLIP-2~\citep{li2023blip}, and Kosmos~\citep{2023MLLM_survey}, are noted in the literature for their ability to assist users in visual-related information.

Video information is gaining prominence in vision assistance, and \citet{Zhao2023LearningVideo} has introduced a novel approach to automatically narrate lengthy videos using LLMs. UniViLM is another language pre-trained model designed for both multimodal understanding and generation. It is capable of retrieving a video segment based on text descriptions, generating captions for given video clips, segmenting a video according to a provided text input, and performing multimodal sentiment analysis of a video segment~\cite{luo2020univl}. VidIL and LLaViLo are additional MLLMs with similar capabilities, demonstrating proficiency in video classification and video-language operations such as video captioning, video question answering, video caption retrieval, and prediction of future video events~\cite{Wang2022LanguageFewShoT,qian2022multimodal}.

These MLLMs adhere to a shared task set, that includes visual question answering, visual captioning, visual common-sense reasoning, visual generation, multimodal affective computing, visual retrieval, vision language navigation, multimodal machine translation, visual question generation and visual dialoguing, as summarised in Table~\ref{tab:MLLM_capabilities}~\cite{Kiros2014MultimodalLM,Shagun2022MultimodalReviw}.

\begin{table*}[]
\centering
\caption{Capabilities of MLLMs trained for vision assistance}\label{tab:MLLM_capabilities}
\begin{tabular}{llccccccccccc}
\textbf{Capabilities}    &       & \rotatebox{90}{\textbf{GILL}} & \rotatebox{90}{\textbf{VisionLLM}} & \rotatebox{90}{\textbf{GPT-4}} & \rotatebox{90}{\textbf{LLaVa}} & \rotatebox{90}{\textbf{MiniGPT-4}} & \rotatebox{90}{\textbf{\parbox{2cm}{Visual\\ ChatGPT}}} & \rotatebox{90}{\textbf{\parbox{2cm}{InternGPT \\with Husky}}} & \rotatebox{90}{\textbf{Flamingo}} & \rotatebox{90}{\textbf{Kosmos}} & \rotatebox{90}{\textbf{UniVL}} & \rotatebox{90}{\textbf{VidIL}} \\ \midrule
\rowcolor[HTML]{EFEFEF} 
\cellcolor[HTML]{EFEFEF}                                                                                                                                                           & Image & $\checkmark$  & $\checkmark$       & $\checkmark$   & $\checkmark$   & $\checkmark$       & $\checkmark$                                                       & $\checkmark$                                                             & $\checkmark$      & $\checkmark$    &                &                \\
\rowcolor[HTML]{EFEFEF} 
\multirow{-2}{*}{\cellcolor[HTML]{EFEFEF}\begin{tabular}[c]{@{}l@{}}Visual question answering\\ {\tiny(Task of providing an answer to a visual input.)}\end{tabular}} & Video &               &                    &                &                &                    &                                                                    &                                                                          &                   &                 & $\checkmark$   & $\checkmark$   \\
                                                                                                                                                                                   & Image & $\checkmark$  & $\checkmark$       & $\checkmark$   & $\checkmark$   & $\checkmark$       & $\checkmark$                                                       & $\checkmark$                                                             & $\checkmark$      & $\checkmark$    &                &                \\
\multirow{-2}{*}{\begin{tabular}[c]{@{}l@{}}Visual captioning\\ {\tiny(Task of generate fitting visual descriptions.)}\end{tabular}}                                                      & Video &               &                    &                &                &                    &                                                                    &                                                                          &                   &                 & $\checkmark$   & $\checkmark$   \\
\rowcolor[HTML]{EFEFEF} 
\cellcolor[HTML]{EFEFEF}                                                                                                                                                           & Image & $\checkmark$  & $\checkmark$       & $\checkmark$   & $\checkmark$   & $\checkmark$       & $\checkmark$                                                       & $\checkmark$                                                             & $\checkmark$      & $\checkmark$    &                &                \\
\rowcolor[HTML]{EFEFEF} 
\multirow{-2}{*}{\cellcolor[HTML]{EFEFEF}\begin{tabular}[c]{@{}l@{}}Visual common-sense reasoning\\ {\tiny(Task of infer understanding from images or video clip.)}\end{tabular}}         & Video &               &                    &                &                &                    &                                                                    &                                                                          &                   &                 & $\checkmark$   & $\checkmark$   \\
                                                                                                                                                                                   & Image & $\checkmark$  &                    &                &                &                    & $\checkmark$                                                       & $\checkmark$                                                             &                   &                 &                &                \\
\multirow{-2}{*}{\begin{tabular}[c]{@{}l@{}}Visual generation\\ {\tiny(Task of generating image or video from a given textual input.)} \end{tabular}}                                      & Video &               &                    &                &                &                    &                                                                    &                                                                          &                   &                 &                &                \\
\rowcolor[HTML]{EFEFEF} 
\cellcolor[HTML]{EFEFEF}                                                                                                                                                           & Image & $\checkmark$  & $\checkmark$       & $\checkmark$   & $\checkmark$   & $\checkmark$       & $\checkmark$                                                       & $\checkmark$                                                             & $\checkmark$      & $\checkmark$    &                &                \\
\rowcolor[HTML]{EFEFEF} 
\multirow{-2}{*}{\cellcolor[HTML]{EFEFEF}\begin{tabular}[c]{@{}l@{}}Multimodal affective computing\\ {\tiny(Task of automatically recognition of emotions and causes.)}\end{tabular}}     & Video &               &                    &                &                &                    &                                                                    &                                                                          &                   &                 &                &                \\
                                                                                                                                                                                   & Image & $\checkmark$  & $\checkmark$       &                &                &                    & $\checkmark$                                                       & $\checkmark$                                                             &                   &                 &                &                \\
\multirow{-2}{*}{\begin{tabular}[c]{@{}l@{}}Visual retrieval\\ {\tiny(Task of language and vision understanding and retrieval.)}\end{tabular}}                                            & Video &               &                    &                &                &                    &                                                                    &                                                                          &                   &                 & $\checkmark$   &                \\
\rowcolor[HTML]{EFEFEF} 
\cellcolor[HTML]{EFEFEF}                                                                                                                                                           & Image & $\checkmark$  &                    &                &                &                    & $\checkmark$                                                       & $\checkmark$                                                             &                   &                 &                &                \\
\rowcolor[HTML]{EFEFEF} 
\multirow{-2}{*}{\cellcolor[HTML]{EFEFEF}\begin{tabular}[c]{@{}l@{}}Vision-language navigation\\ {\tiny(Task of navigation based on linguistic instructions.)}\end{tabular}}              & Video &               &                    &                &                &                    &                                                                    &                                                                          &                   &                 & $\checkmark$   &                \\
                                                                                                                                                                                   & Image &               & $\checkmark$       & $\checkmark$   & $\checkmark$   & $\checkmark$       & $\checkmark$                                                       & $\checkmark$                                                             & $\checkmark$      & $\checkmark$    &                &                \\
\multirow{-2}{*}{\begin{tabular}[c]{@{}l@{}}Multimodal machine translation\\ {\tiny(Task of translation from a video or an image)}\end{tabular}}                                          & Video &               &                    &                &                &                    &                                                                    &                                                                          &                   &                 &                &                \\
\rowcolor[HTML]{EFEFEF} 
\cellcolor[HTML]{EFEFEF}                                                                                                                                                           & Image &               & $\checkmark$       & $\checkmark$   & $\checkmark$   & $\checkmark$       & $\checkmark$                                                       & $\checkmark$                                                             & $\checkmark$      & $\checkmark$    &                &                \\
\rowcolor[HTML]{EFEFEF} 
\multirow{-2}{*}{\cellcolor[HTML]{EFEFEF}\begin{tabular}[c]{@{}l@{}}Visual question generation\\ {\tiny(Task of generating questions for given image or video)}\end{tabular}}             & Video &               &                    &                &                &                    &                                                                    &                                                                          &                   &                 &                & $\checkmark$   \\
                                                                                                                                                                                   & Image & $\checkmark$  &                    & $\checkmark$   & $\checkmark$   & $\checkmark$       & $\checkmark$                                                       & $\checkmark$                                                             & $\checkmark$      & $\checkmark$    &                &                \\
\multirow{-2}{*}{\begin{tabular}[c]{@{}l@{}}Visual dialoguing\\ {\tiny(Task of  automating a conversation about a video or image)}\end{tabular}}                                          & Video &               &                    &                &                &                    &                                                                    &                                                                          &                   &                 &                & $\checkmark$  
\end{tabular}

\end{table*}

\subsection{Conversation}

Specific LLMs are trained explicitly for meaningful and coherent dialogues with humans. An example is Dialogue Generative Pre-trained Transformer (DialoGPT), a fine-tuned model trained on 174 million Reddit conversations~\cite{2020ZhangDIALOGPT}. DialoGPT exhibits the ability to provide human-like answers in tested conversations~\cite{2020ZhangDIALOGPT}. \citet{dettmers2023qlora} introduced a fine-tuning mechanism for LLMs named Quantised Pre-trained Language Model into Low-Rank Adapters (QLoRA). This allows for the fine-tuning of large-parameter LLMs with low training costs. They introduced Guanaco, a fine-tuned LLM with 65 billion parameters, which achieved a performance level of 99.3\%. Falcon-180B and Falcon-40B represent another set of open-source LLMs with 180 billion and 40 billion parameters. These models are trained to communicate in multiple languages, allowing users to engage in conversations in languages other than English~\cite{2023GuilhermeFalcon}. To evaluate the accuracy of human-like dialogue systems, \citet{ou2023dialogbench} proposed a dialogue evaluation benchmark named DialogBench, which consists of 12 dialogue tasks to assess the capabilities of LLMs. In their evaluation, they assessed 28 pre-trained and instruction-tuned LLMs, demonstrating that GPT-4, ChatGPT, and KwaiYii-13B-Chat emerged as the top three models for conversations in domains related to daily life and professional knowledge. 

In DF chat conversations, the significance lies in facilitating non-technical investigators to elucidate terminologies and areas lacking understanding. This serves a dual purpose, acting as an interactive teacher to enhance comprehension in discussions~\cite{2023SCANLONChatGPT}.

\subsection{Prompt Engineering}
Achieving quality outputs from LLMs often relies on providing well-crafted, meaningful, and precise input queries, known as input prompts. However, even human-defined natural language instructions may not consistently yield the best results. Prompt engineering is a methodology that involves carefully defining and instructing LLMs to generate more accurate and desirable outputs. Through thoughtful refinement of input prompts, prompt engineering aims to enhance the performance and effectiveness of LLMs in generating output that align more closely with user expectations and requirements~\cite{marvin2024Prompt, zhou2022large}. This plays a crucial role in biasing LLMs toward specific domains or topics, enabling a more targeted and nuanced response. By carefully crafting prompts, users can guide LLMs to dive deeper into the nuances of their queries, leading to more accurate and relevant outputs. This approach enhances the model's responsiveness to specific areas of interest, allowing users to fine-tune and tailor their interactions with the LLM for more precise and meaningful outcomes. Prompt engineering with LLMs is employed in various sectors, including, but not limited to, medical, engineering, construction, and healthcare~\cite{2022MeskoPrmotMedical,polak2023extracting}.

ChainForge is an open source Graphical User Interface (GUI) tool developed specifically for prompt engineering and hypothesis testing derived from LLMs that can be used in the aforementioned fields to generate accurate and quick output~\cite{arawjo2023chainforge}.

Although prompt engineering is a necessity in generating the desired output from an LLM, the results can still be biassed based on the wording and phrasing provided by the user. Since the effectiveness of prompts depends on the user's proficiency in English, the output may vary significantly depending on the exact requirements of the user and how the LLM interprets these prompts. In addition, the prompts can unintentionally reinforce existing biases within the model's training data, potentially skewing the results. Therefore, prompt engineering must be approached carefully and methodically to minimise misinterpretation and maximise output relevance.

\subsection{Autonomous Agents}
\label{AI_agents}

The evolution of LLMs, with their ability to generate information and communicate in a manner that resembles human interaction, has led to the development of autonomous agents. The expectation is that these agents will effectively perform a wide range of tasks, taking advantage of the human-like capabilities inherent in LLMs~\cite{wang2023survey}. These autonomous agents follow a four-stage architecture that includes profiling, memory, planning, and action. Profiling defines the agent's role, privileges, domain, and expertise~\cite{wang2023survey}. Memory stores information on tasks and profile data relevant to the environment. Planning involves breaking down given tasks into subtasks and solving them individually. The action stage is the final phase where all decisions and subtasks are translated into actions executed by the agent.
\citet{zhang2023building} developed a framework designed to facilitate collaboration between GenAI agents and humans. This framework enables planning and communication for specific tasks, leveraging the capabilities of LLMs Similarly, AgentSims~\citep{lin2023agentsims}, ToolBench~\citep{qin2023toolllm}, GameGPT~\citep{chen2023gamegpt}, ChatDev~\citep{qian2023communicative}, Voyager~\citep{Qin2024Voyager}, and RecMind~\citep{wang2023recmind} represent a diverse array of autonomous GenAI agents developed with distinct goals and objectives. Certainly, AutoGen stands out as a multiagent framework with the capability to autonomously perform tasks or collaborate with human feedback. This flexibility makes it a versatile tool for various applications~\cite{wu2023autogen}.

In addition to the AutoGen framework, AgentLite~\citep{liu2024agentlite}, Camel~\citep{li2023camel} and CrewAI~\citep{10.1007/978-3-031-70415-4_4} are each similar LLM-based agent framework architectures. These platforms are distinguished by their support for task decomposition, multi-agent orchestration, and adaptable reasoning. In particular, AgentLite and CrewAI facilitate work delegation functionalities, increasing their utility in various operational contexts.

\subsection{Retrieval-augmented Generation}

The content generated by LLMs is highly dependent on the extensive text-based datasets on which they are trained. These datasets may contain vast amounts of information utilised by the LLMs. However, maintaining up-to-date knowledge within LLMs is challenging, as fine-tuning or retraining a model is often extremely costly and resource-intensive. Retrieval-augmented Generation (RAG) is a technique designed to address this LLM knowledge gap by retrieving information from external sources and integrating it with the model's internal representations~\cite{Fan2024SurveyRAG}.

A major advantage of RAG is its ability to reduce the hallucination problem in LLMs, allowing them to generate more accurate and current information~\cite{jiang2023active, lewis2020RAG}. The architecture of RAG consists of a knowledge base and a retriever model. The retriever model converts input prompts and content from the knowledge base to vectors. The user prompts are then appended with the most relevant content from the knowledge base, and this augmented prompt is sent to the base model to generate a more accurate response~\cite{lewis2020RAG}.

\subsection{Limitations and Risks}
\label{limitations}

As explored in the preceding sections, LLMs appear to possess a vast range of capabilities. However, it is crucial to acknowledge that they are not without limitations and risks. In multimodal LLMs, it is a common problem that they are over-reliant~\cite{subra2024CognitiveChallenges}. There are also potentially significant drawbacks associated with LLMs, including issues such as bias, explainability challenges, reasoning errors, logical errors, hallucinations, vulnerability to prompt injections, and spelling and grammar errors. These limitations underscore the importance of a cautious and critical approach when using LLMs in various applications. Furthermore, the literature shows limitations in LLMs, including statistical inconsistency, the absence of emotional attributes in linguistic responses, and challenges related to fact verification~\cite{tang2023science,frohling2021feature}. 
These factors contribute to a comprehensive understanding of the constraints and potential shortcomings when working with LLMs. \citet{thapa2023humans} contend that while LLMs can indeed reduce the time and costs associated with annotation tasks, they are not completely supplanting human annotation. This is because they struggle with intricate linguistic constructions, such as idioms, irony, sarcasm, and metaphor, which can potentially impact the precision of annotations.

Similar limitations are associated with MLLMs, such as over-reliance on training data, sensitivity to word order in input prompts, and vulnerability to prompts containing additional knowledge~\cite{Shuhan2023MLLMsIssues}. There are several more concerns associated with LLMs, including restricted text input and output lengths, limited comprehension of syntax, ethical considerations with the generated information, constraints with multilingual capabilities, elevated costs associated with training and maintenance, inadequate understanding of human behaviours and limited ability to learn incrementally \cite{chiang2023large,Alawida2023ChatGPTLimits}.

Despite their considerable capabilities, LLMs are not without risks. \citet{bommasani2021opportunities,lund2023chatting, rahman2023beyond} provide comprehensive overviews of risks linked to LLMs. These include the homogenisation of results, whereby defects or biases from the foundation model are inherited by all downstream models.   There is also the risk of monopolistic control by foundation model owners, potentially concentrating decision-making power, resource access, and influence over model usage in a single entity.  Ethical and legal concerns are intertwined with concerns about privacy and intellectual property. In addition, there are economic and environmental impacts that raise concerns about the potential displacement of human workers. Furthermore, inequity and misuse of LLMs, such as the creation of deepfakes and their application in criminal and unethical activities, pose additional challenges. Given that LLMs do not inherently prioritise the precision of information, \citet{2021BenderOntheDanger} have highlighted the risk of generating social turbulence, especially when used on social media platforms. Furthermore, the use of LLMs is associated with significant costs, leading to a direct environmental impact due to their substantial energy consumption~\cite{2023RilligRisks}.

The risks associated with LLMs are predominantly emphasised within Information Communication and Technology (ICT) and cyberspace. Primary concerns include the disclosure of personal information, the generation of malicious text, and the creation of malicious code~\cite{rao2023ethical}.

The Beyond the Imitation Game benchmark (BIG-bench), serves as an evaluation framework for LLMs. It encompasses 204 distinct language-related tasks. These span contextual and context-free question-answering, reading comprehension, logical reasoning, etc.~\cite{srivastava2022beyond}. It is acknowledged that the challenge of social biases and dependency on the English language persists in almost all LLMs.

When employing LLM-based agents, it is imperative to address challenges associated with LLM-based multi-agent frameworks as well. The handling of many defined agents may necessitate substantial computational resources and memory, thus mandating high-end computing infrastructure for seamless operations. Furthermore, the absence of a standardised comprehensive benchmarking system to evaluate the behaviour of such agents underscores the limitations inherent in the development of LLM-based multi-agent systems. These challenges underscore the need for further research and refinement in this domain to enhance the efficiency and effectiveness of LLM-based multi-agent frameworks~\cite{guo2024large}.  

Similarly to these risks, the Open Web Application Security Project (OWASP) has identified ten major risk factors related to LLMs. These risks include training data poisoning, prompt injection, denial of service, insecure output handling,  supply chain vulnerabilities, sensitive information leakage, excessive agency, insecure plugins, overreliance, and model theft of data. Despite these threat factors, OWASP also stressed the need for regulatory bodies to supervise LLMs in various domains and recommended the implementation of risk management programmes that incorporate the checklist provided by OWASP\footnote{https://owasp.org/www-project-top-10-for-large-language-model-applications}.

The EU's Artificial Intelligence Act (AIA) proposes a framework for categorising AI applications based on their associated risk levels, with the primary aim of safeguarding human rights and maintaining ethical standards in AI deployment~\cite{NEUWIRTH2023105798}. Within the AIA, AI applications are divided into categories such as ``unacceptable risk'', which includes practices such as exploiting vulnerabilities and social-ranking techniques due to their potential for individual manipulation and impact on fundamental rights. These categories have relevance to DF, where the ethical application of LLMs must balance the advantages of automation with the imperative to address privacy concerns. Since DF involves sensitive data and influences legal outcomes, the use of LLMs must align with AIA's risk-based principles, ensuring transparency, accountability, and fair application. In future AI applications within DF, it is essential to implement appropriate measures to prevent biases and hallucinations to mitigate the risk of misuse of AI.

%

%

\section{Large Language Models For Digital Forensics}
\label{LLM_for_DF}

Section~\ref{LLM_for_DF} summarises existing work with LLMs in DF, the feasibility of employing them, and potential future directions. As discussed in Sections~\ref{LLMS} and \ref{LLMS_capabilities}, despite the widespread use of LLMs in various fields to improve the efficiency and accuracy of tasks within specific domains, their application in the field of DF is still relatively new.


Conducting a thorough analysis of LLMs use in conjunction with the stages of the DF process model, as highlighted in Section~\ref{DF_process_models}, proves to be a valuable undertaking.

\subsection{Incident Recognition Phase}
In the initial phase of Casey's DF process model, which delineates the recognition of an incident, LLMs can serve as a valuable detection mechanism~\cite{goel2024xlifecycle}. In cybercrime cases, the primary artefacts often involve data logs, data dumps and network dumps. Fine-tuning an LLM to monitor text-based logs and related files enables it to discern and identify potential or ongoing incidents within the environment.
In network-related activities, anomaly detection plays a pivotal role in initiating an incident response. Various existing anomaly detection techniques are employed in systems for this purpose. Using their ability to identify patterns in a series of text data sets, LLMs exhibit potential as an Intrusion Detection System (IDS) within such systems~\cite{li2023myriad,yang2024AnomalyDetect}.
  
For instance, \citet{kan2024Mobile-LLaMA} introduces Mobile-LLaMA, an open source mobile network-specialised LLM, fine-tuned through instructional data to enhance their capabilities for network analysis tasks within 5G environments. Mobile-LLaMA supports three primary functions: IP routing analysis, packet analysis, and performance analysis.

\subsection{Collection Phase}


Although evidence collection or seizure traditionally involves physical tasks that require human interaction, LLMs can play a role in identifying and listing potential pieces of evidence at a crime scene. For example, in the examination of photographs or video records from a crime scene, an investigator can enlist the help of a MLLM such as LLaVa, GPT-4, or VisionLLM. These models are capable of processing information within the images and generating a text-based output, facilitating the interpretation and categorisation of visual data. Although this task may seem simple and within the capabilities of a human agent, the efficiency becomes particularly evident when dealing with a massive-scale investigation involving thousands of collected artefacts and photographs. Using an MLLM for initial processing can significantly save time, with human agents then focussing on the crucial task of verification and validation.


\subsection{Preservation/Acquisition Phase}

The preservation of evidence is centred on maintaining integrity. To achieve this, various tools such as EnCase and FTK Imager have been used, helping investigators streamline their work processes~\cite{shah2017protecting}. In the context of non-technical DF stakeholders being able to interrogate the evidence, it becomes feasible for a user to articulate their requirements/query in natural language. Subsequently, the LLM generates source code tailored to the specific need, executes the code on the data, and returns the result in consumable natural language.
  

LLMs specialised in code generation, such as StarCoder and Code LLaMA, can be fine-tuned and retrained for domain-specific tasks, including the preservation of disk evidence through customised code and script generation. These LLMs are capable of generating scripts or code snippets that create secure copies of disk images, metadata, and partition information, as well as automating cryptographic hashing and verification routines to maintain the evidence’s integrity through checksums. Additionally, LLMs can assist in documenting preservation steps by generating logs and summaries for each stage of the disk preservation process, thereby supporting the chain of custody during acquisitions. However, despite these capabilities, human expertise remains essential for identifying and collecting potential sources of evidence during the preservation phase, as the application of LLMs in this stage is currently limited to lower-potential tasks.

In certain instances, the gathering of live data for forensic investigations becomes crucial, particularly data collected at the crime scene. For this purpose, investigators can use DFaaS platforms such as Hansken. Hansken possesses the ability to amalgamate custom extraction APIs for data extractions, and these APIs can be developed using code-generative LLMs~\cite{2015VANBEEKDFaaS}. This approach improves the adaptability and efficiency of the investigative process.

As stated in Section~\ref{AI_agents}, the automation of code generation and unit testing can be facilitated by autonomous agents that use LLMs as their core. AutoGen, being an open source framework, provides the means to develop AI agents tailored for specific tasks. These AutoGen agents are not only customisable and conversational, but can also operate in various modes, employing combinations of LLMs, human inputs, and various tools~\cite{wu2023autogen}. 
  
Automated agents, particularly those developed within frameworks such as AutoGen, can be used in the preservation phase of investigations. These agents can be assigned specific tasks, such as acquiring disk images, generating disk hashes, retrieving disk metadata, and compiling acquisition reports. By defining precise roles and tasks for AI agents, it is possible to streamline and standardise these preservation actions, improving the management of digital evidence~\cite{wu2023autogen,Akila2024aiagents}.

\subsection{Examination Phase}

This phase constitutes a crucial component of the investigation, playing a crucial role in elucidating the case through activities such as data recovery, collection, reduction, and classification. For each of these components, LLMs fine-tuned for scripting can significantly assist, especially at a larger scale. Within these components, tasks such as keyword search, file recovery, pattern matching, and fragment reassembly can be achieved with minimal technical knowledge using LLMs. LLMs can provide valuable assistance in these tasks by generating new codes, crafting regular expressions, generating passwords and/or password hash lists for decryption, and creating sample logs or files. 
LLMs can generate a set of instructions, queries, and Application Programming Interface (API) validations from natural language provided by a human. This opens up the possibility of integrating third-party tools like Scapy, tshark, John the Ripper, and others seamlessly into the investigative process, enhancing the toolkit available for DF investigations, and the ability to automate these processes enhances efficiency and effectiveness in the examination phase of the investigation.  

The use of LAMs and VoT techniques in the examination phase can significantly enhance the efficiency of an investigation. Since LAMs and VoT specifically focus on task manipulation, investigators can offload some examination work to an LAM, which will then generate the final results from a series of subtasks. This approach can allow investigators to focus on higher-level analysis and decision making, thus streamlining the overall investigative process.

\subsection{Analysis Phase}
The analysis phase involves understanding the incident and obtaining a conclusive understanding based on the information collected during the examination phase. As also highlighted in Section~\ref{exsisting_workDF}, it has been demonstrated that LLMs are effective in case analysis~\cite{henseler2023chatgpt}. The use of MLLMs, which possess the capability to interpret images, broadens the scope for analysing a crime case more comprehensively.
 
Using Gemini 1.5, \citet{10.1145/3627673.3679091} presented a tutorial on profiling a suspect's web history through an LLM. This case study demonstrates how an LLM can help identify potential motivations, personal interests, and psychological characteristics of the suspect. In conclusion, the authors suggest that such mechanisms could power AI-assisted tools, enabling law enforcement authorities to improve the identification of cybercriminals and malicious entities.

The Digital Forensic Cybercrime Language as a Service (DFClaaS) is an innovative system developed to address the complexities of text-based cybercrime~\cite{al2024proof}. Using natural NLP techniques, including LLMs, sentiment analysis, and lexicon analysis, DFClaaS aims to improve capabilities in incident reporting, analysis, and investigation. The primary objectives of DFClaaS include implementing microservices to address specific challenges, proposing an advanced system to improve incident handling, and providing valuable tools for DF investigators. Designed to serve individual users, organisations, and forensic professionals, DFClaaS is a versatile and effective resource in the ongoing fight against cybercrime.

LLMs can be specifically fine-tuned for the analysis of various data types, including log files, email contents, chat transcripts, call records, file metadata, hex dumps, memory dumps, and registry hives. Incorporating contents such as event logs, timestamps, and network traffic captures further enables the effective recreation of incidents by correlating each data set with the assistance of LLMs. In addition, MLLMs that are audio and video specific can assist in analysing content within these formats. This specialised capability can significantly reduce the time investigators spend analysing audio and video data during investigations.

The use of automated agents can effectively distribute the analysis workload. Moreover, leveraging Augmented Large Language Models (ALLMs) and RAG techniques can improve knowledge retrieval in real time, thus improving the accuracy of analysis and decision-making processes~\cite{Lewis2020AgLLM}. For example, integrating a source of intelligence with an RAG system can assist investigators in connecting the dots during a DF investigation.

Other than these applications, LLMs can increase productivity through enhanced information correlation during the analysis phase. \citet{SHAFEE2025125509} suggest that LLMs hold significant potential for data correlation and cybersecurity applications. The referenced study evaluated the performance of various LLM-based chatbots, including ChatGPT, GPT4all, Dolly, Stanford Alpaca, Alpaca-LoRA, Falcon, and Vicuna, specifically for text classification and Named Entity Recognition (NER) tasks using OSINT data. The findings indicated that, although the commercial chatbot GPT-4 and the open-source GPT4all performed well in text classification, all tested LLM-based chatbots showed limitations and were less effective for cybersecurity entity recognition compared to specialised models. The study concludes that there remains room for improvement.

\subsection{Reporting Phase}

The quality and validity of the evidence, along with the thoroughness of the analysis, are encapsulated in the final report. The reporting phase holds significant weight, as the entire judgement may hinge on this crucial stage. Notably, DF is experiencing heightened scrutiny about the quality of the reports, emphasising the importance of precision and clarity in this phase~\cite{2019NicksonStandardising}. As pointed out by \citet{champod2016enfsi}, there is no standard framework for evaluating and reporting scientific findings to authorities and stakeholders. 
To provide assistance and alleviate scrutiny, incorporating LLMs for report creation is a viable solution. 
 
While LLMs are inherently non-deterministic, adhering to investigation standards such as ISO/IEC 27043:2015 can establish robust processes around data integrity and evidence handling, even though these standards do not directly address the randomness or variability in LLM outputs. The ISO/IEC 27043:2015 standard provides guidelines for a consistent DF investigation framework, focusing on maintaining procedural rigour rather than modifying model behaviour. Although it does not directly resolve issues of LLM determinism, it can serve as a protocol to ensure that procedures involving LLMs uphold investigative standards and maintain integrity throughout the process~\cite{2016ValjarevicISO}. 

  A preliminary feasibility study by~\citet{michelet2023chatgpt} highlighted the potential of LLMs to assist in automating forensic report generation. These models can facilitate the creation of structured sections, such as methodologies, data analysis, and summaries, by generating coherent, case-specific insights from forensic data. Additionally, LLMs could automate the production of reports in alternative formats, such as HTML or \LaTeX, which are frequently used for dynamic, web-based, or highly technical documentation.


\begin{table*}[h] 
    \caption{DF functionalities by CFTT highlighting the usability of LLMs and example prompts}\label{tab:DF_example_prompts}
    \begin{tabular*}{\linewidth}{p{0.18\linewidth} p{0.1\linewidth} p{0.29\linewidth} p{0.35\linewidth}}
        \toprule
\textbf{CFTT Functionality} & \textbf{DF Phase(s)}        & \textbf{Usable LLMs/Agent Frameworks}                                        & \textbf{Example Prompt}                        \\
        \midrule
Cloud Data Extraction                & Acquisition              & LLaMA (Fine-Tuned), Code Llama, StarCoder, AutoGen or CrewAI                     & Retrieve all the data inside given S3 bucket by using given credentials         \\ \hline
Deleted File Recovery Specs          & Acquisition, Examination & LLaMA (Fine-Tuned), Code Llama, StarCoder, AutoGen or CrewAI                     & Find all the deleted files from the X disk image and recover them to Y location \\ \hline
Disk Imaging                         & Acquisition              & LLaMA (Fine-Tuned), Code Llama, StarCoder, AutoGen or CrewAI                     & Get a full disk image from this computer and save it in Z location             \\ \hline
Forensic File Carving                & Acquisition, Examination & LLaMA (Fine-Tuned), Code Llama, StarCoder, AutoGen or CrewAI                     & Find all the deleted PDF files from X disk image                            \\ \hline
Forensic Media Preparation           & Acquisition              & LLaMA (Fine-Tuned), Code Llama, StarCoder, AutoGen or CrewAI                     & Prepare the given X device for new investigation                                \\ \hline
String Search                        & Examination              & LLaMA (Fine-Tuned), Code Llama, StarCoder, AutoGen or CrewAI                     & Search all the files containing the email of mail@test.com                     \\ \hline
Mobile Devices                       & Examination              & LLaMA (Fine-Tuned), LLaVa (Fine-Tuned), Code Llama, StarCoder, AutoGen or CrewAI & Find all the photos taken with a computer within the last 3 months             \\ \hline
MS Windows Registry                  & Examination              & LLaMA (Fine-Tuned), Code Llama, StarCoder, AutoGen or CrewAI                     & Find information about the users from a given Windows disk image                \\ \hline
SQLite Databases                     & Examination              & LLaMA (Fine-Tuned), Code Llama, StarCoder, AutoGen or CrewAI                     & Find the access time of user Y to application X using given SQLite databases \\
        \bottomrule
    \end{tabular*}
\end{table*}

\subsection{Other Possibilities}


\citet{2023SCANLONChatGPT} highlights that LLMs can play an important role in teaching scenarios. This involvement extends to activities such as storyboarding, creation of synthetic content, and synthetic character profiling. 
  
Fine-tuned models could further enhance training by generating more complex, realistic case examples that challenge trainees with nuanced scenarios, providing a robust foundation for practical skills development. These models may also help translate technical findings into accessible language, facilitating communication of insights to non-specialists, such as judges or other stakeholders.

\subsection{Discussion on Potential for LLMs in DF}
To provide a comprehensive understanding of the potential use of LLMs, Table~\ref{tab:DF_example_prompts} clarifies the sample functionalities within the framework of the National Institute of Standards and Technology (NIST) Computer Forensics Tool Testing Program (CFTT), highlighting the usability of LLMs and example prompts. The CFTT project establishes overarching specifications to assess the capabilities of tools, a framework adopted by numerous prominent free and commercial tools~\footnote{https://www.nist.gov/itl/ssd/software-quality-group/computer-forensics-tool-testing-program-cftt}.

The potential for having a positive impact on the typical phases of the investigation increases as one progresses through the typical order of the phases. For example, there is little improvement that can be made by an LLM or automated scripting during the identification or acquisition phases, but significant potential for aiding investigators in the reporting phase~\cite{michelet2023chatgpt,Akila2024aiagents} -- these are first and foremost large \textit{language} models. The low/medium/high potential outlined below evaluates each DF phase based on three key requirements: reliance on human expertise, physical versus digital evidence handling, and scope for automation, as explained below.

\begin{itemize}
    \item \textbf{Low Potential for the Identification and Collection Phases}
        \begin{itemize}
            \item High dependency on human involvement, expertise, and/or specialised knowledge.
            \item Involves extensive handling of physical evidence.
            \item Limited or no feasibility for automation.
        \end{itemize}
    \item \textbf{Medium Potential for the Preservation Phase}
        \begin{itemize}
            \item Requires some level of human involvement or expertise, but is not critical to the process.
            \item Primarily deals with digital evidence, with minimal physical evidence handling.
            \item Feasible for automation to a significant extent.
        \end{itemize}
    \item \textbf{High Potential for the Examination, Analysis and Reporting Phases}
        \begin{itemize}
            \item Human involvement is needed for expert verification of the conducted analysis.
            \item Exclusively focused on digital evidence.
            \item Many common tasks are suitable for significant support from LLMs.
        \end{itemize}
\end{itemize}

With these possibilities, the scope for research in DF is vast. Future research could be extended to the generation of digital forensic reports, as well as the summarisation of these reports for non-technical users. This would save time, but can also lead to more consistent documentation compared to manual documentation. Given the capacity of LLMs to manage large textual datasets, exploring pattern recognition holds significant value, particularly for investigations requiring the detection of anomalies or outliers in chats, log events, or emails. 

In addition, LLMs’ ability to interpret the tone of messages or chats enables their application in the sentiment analysis of text-based evidence. There is also potential in fine-tuning LLMs for domain-specific tasks, such as network forensics, where LLMs could analyse log files and application data related to specific activities. Automating LLM-based DF tools could further enable investigators to generate customised reports using natural language queries. 

A critical future research direction lies in the ethical and legal considerations of LLM-generated content. As the application of LLMs is still emerging, future studies should focus on developing appropriate benchmarks, standardisation protocols, and addressing legal aspects to ensure responsible use of this technology.

\section{Challenges and Risks}

This section discusses the challenges and risks of using LLMs in DF. Despite their promising potential, there are significant risk factors to consider. These risks can have severe consequences for DF if not adequately identified and considered in the DF process.

\subsection{Challenges for LLMs in Digital Forensics}
To optimise the results, the LLMs will likely need to be trained with specific forensic data (i.e., previous case data) to achieve the best results. Given the complexity and variation of the cases, it is questionable how good the training data are and whether there are sufficient data~\cite{breitinger202410YearReview}. Any bias in training data can lead to skewed interpretations and unjust outcomes. 
  This problem of bias can be mitigated by using diverse and representative datasets during the training phase, e.g., datasets that come from diverse sources, different case types and geographic regions. Furthermore, techniques such as data filtering, distribution reconstruction, rebalancing, regularisation, and prompting can be implemented to actively identify and correct biased patterns in the base data sets of the model and its outputs~\cite{dai2024BiasMitigation,Zhou_2024LLMBias}. These techniques involve adjusting model weights or incorporating fairness constraints during training to reduce the likelihood of biased predictions. Regular audit of the model's performance against fairness benchmarks is also crucial to ensure that it remains unbiased over time~\cite{mokander2023auditing}.

The experience level of investigators and the practical strategies employed in conducting investigations are challenging to replicate with LLMs. 
   Initially, LLMs can excel in assisting with certain subtasks, such as parsing and data conversion, tasks in which output can be easily verified. However, when it comes to more interpretative or inferential tasks, LLMs’ lack of inherent transparency introduces explainability challenges. Unlike deterministic software, whose logic can be easily traced, LLMs often act as black boxes, making it difficult to validate and understand the rationale behind their conclusions, particularly when these outputs extend beyond straightforward parsing into areas requiring judgement and reasoning. This underscores the importance of explainability in the application of LLMs to forensics, where understanding the basis of each result is crucial for accuracy and accountability~\cite{michelet2023chatgpt}.

Publicly hosted and maintained LLMs are generally unsuitable for casework due to the sensitivity of the evidence and information involved, which require strict privacy and security controls that cannot be reliably ensured on public platforms. Furthermore, managing the substantial infrastructure needed for LLM training and deployment is both energy and resource intensive, presenting a financial hurdle, especially for smaller forensic laboratories with limited budgets. Although methods like retrieval-augmented generation (RAG) or prompt engineering can reduce some of the computational load by tailoring responses with existing models, they still require powerful GPU resources to effectively run these models, adding to the cost and accessibility barriers. Centralised systems could be a viable option, but they require well-defined guidelines for data sharing and stringent security standards to safeguard sensitive information.

Although LLMs can serve as valuable tools to support forensic investigations, it should be recognised that they currently function best as an aid, not a substitute for human expertise~\cite{scanlon2023ChatGPT}. There is a risk that people may place too much trust in the results generated by LLMs (over-reliance), which could lead to complacency and overlook the need for detailed human expert analysis and validation. 

To mitigate the potential misuse of LLMs, many LLMs are subjected to censorship~\cite{Yifan2024LLMSecurity,glukhov2023llm}. Although this censorship may serve as a preventive measure against unethical use, it can pose challenges in the field of DF. For example, if an investigator seeks evidence related to `drugs' or evidence of other illegal material, censorship of the LLM may restrict access to accurate information related to the investigator's query. This limitation underscores the need for a nuanced approach to censorship in LLMs, balancing ethical considerations with the imperative of facilitating effective forensic investigations. In addition, the censorship of generic, publicly accessible LLMs further supports the argument for a discipline-specific DF LLM.

Finally, ethical and legal considerations must also be discussed. Determining accountability in cases where LLMs produce false information or are compromised by hacking. Clarifying responsibilities between developers, users, and regulators is crucial to establish a framework for accountability. If LLM generated DF results lead to incorrect information, the responsibility may lie with both the developers, for ensuring the model's accuracy, and the users, for appropriately interpreting and validating the results.

\subsection{Risks of Integration}
\label{risks}
The integration of LLMs within the DF process 
comes with inherent risks, in addition to the general LLM limitations outlined in Section~\ref{limitations}. In particular, in the examination, analysis, and reporting phases, the use of LLMs introduces the risk of producing inaccurate information, primarily due to the phenomenon of inheritance hallucinations associated with these models~\cite{michelet2023chatgpt,scanlon2023ChatGPT}. Additionally, the biases and obscurities present in an inheritance model can significantly impact the performance of a DF-focused LLM -- potentially leading to the unacceptable generation of biased or inaccurate information within the DF process.

Hallucinations in LLMs present a considerable risk, as they can produce information that appears credible but is incorrect. This can lead law enforcement authorities to form invalid assumptions and make flawed decisions based on unreliable results. 
Additionally, inherent biases in LLMs can influence investigative outcomes, which could affect the fairness and integrity of legal procedures. Data privacy concerns are also prominent, as sensitive information confidentiality may be compromised when using LLMs in DF processes. Together, these factors present substantial challenges to the reliable and ethical application of LLMs.

It is also crucial to acknowledge that DF LLMs, like any complex model, are susceptible to adversarial manipulation~\cite{Zou2023UniversalAT}. This vulnerability poses a substantial risk in the context of sensitive domains such as DF, where the integrity of the information obtained is paramount. Adversarial attacks can compromise the reliability of LLM-generated outputs, potentially influencing the outcomes of various phases within the DF process.

Indeed, despite incorporating human verification, outputs and reports generated by LLMs within DF applications may encounter challenges regarding acceptance within the legal systems of different countries. This highlights a significant usability risk associated with LLM-based DF applications, but one that can be carefully mitigated by limiting the technology's deployment as a human-in-the-loop investigative aid as opposed to directly feeding into any investigative/judicial decision-making processes.

Mitigating these changes and risks can be challenging, particularly in scenarios that involve adopting country-specific legal systems. However, there are potential strategies to address technical challenges such as hallucinations, censorship, and substantial infrastructure costs.

One solution to mitigate hallucination was suggested by~\citet{ji-etal-2023-HallMitigate}, who proposed an interactive self-reflection method for generated knowledge and answers, an approach that has shown promise. Another method of reducing hallucinations is the use of RAG, which provides a larger knowledge base for LLMs to minimise unknown information~\cite{lewis2020RAG}. Other methods such as knowledge graphs, bias detection mechanisms, active learning methods for LLMs, supervised fine-tuning strategies, hallucination mitigation frameworks, and new decoding strategies can also help mitigate hallucinations to some extent~\cite{Perković2024Hallucinations, tonmoy2024comprehensive}.

Censorship issues can be addressed by fine-tuning the model with uncensored information, a technique already applied to the LLaMA and Mistral models, leading to the development of the Dolphin models. An example is the Dolphin-2.0-mistral-7b, which is an uncensored version of the Mistral 7B model~\cite{xu2024safedecoding}.

The high infrastructure costs associated with these models can be mitigated by employing Data Forensics as a Service (DFaaS) platforms such as Hansken. With DFaaS, investigators only need to input queries related to their investigations using personal computers, while the platform manages the model maintenance and computational demands~\cite{2015VANBEEKDFaaS}.

Despite these promising integration risks, the use of LLMs may face limitations in adaptability. The performance of an LLM is inherently tied to the dataset on which it was trained, which means that its ability to respond to new or emerging information is constrained. For example, if an LLM is tasked with identifying possible malware in a system, it may struggle to detect newer malware variants that were not part of its training data~\cite{Yu2024maltrack}. To mitigate such issues, LLMs need to be fine-tuned frequently, which poses its own challenges due to the significant computational power required for such operations.

\section{Conclusion}
\label{conclution}

The convergence of LLMs with an array of technologies represents exciting synergy. Although the utilisation of LLMs in the realm of DF is still in its nascent stages, there is evidence of their substantial potential to significantly increase the efficiency of investigations. The exploration of investments for LLMs across the entire DF process is considered, with the aim of improving the productivity and efficiency of investigations. Additionally, the integration of LLMs into current DF tools is posited to reduce user training times, as these models comprehend natural language input and provide output accordingly. In the dynamic landscape of LLM applications for DF, promising avenues for further exploration and advancement unfold.

Although the surge in LLM research is promising, it is crucial to balance enthusiasm with awareness of existing challenges. The propensity of LLMs to produce hallucinations highlights the need for human oversight in critical decision-making processes, underscoring the irreplaceable value of human judgement, intuition, and expertise. A notable limitation is the language dependency issue, as most LLMs are predominantly trained on English data, reducing their effectiveness with non-English content. Furthermore, the deployment of LLMs in DF involves significant costs related to the infrastructure to process evidence. Questions also arise about the validation of task correctness and quality when automated by LLMs, as well as the legal and professional acceptance of results obtained with limited human intervention.

The trustworthiness of LLMs remains a debatable issue that requires careful attention. It is crucial to establish clear boundaries and measures to define LLM trustworthiness. Addressing this will be a key aspect in the field of DF, ensuring that LLMs can be trusted for accurate and secure analysis, with the explainability of their operations being paramount.



Integrating LLMs with automated agents offers a promising path to automating DF processes, potentially allowing multiple cases to be handled concurrently for more timely and precise outcomes. This integration could significantly streamline investigations. Future research should explore the role of LLMs and AI in the decision making of DF. It is essential to focus on validating LLM generated outputs to ensure their scope, accuracy, reliability, and trustworthiness in investigations. More studies comparing DF outcomes with and without LLM integration are critical, as they could highlight the benefits of LLMs and the controlled applicability of LLMs in DF and similar fields.

A future use case involves developing forensic-specific LLMs fine-tuned for automated examinations. These models could be optimised for script generation to support investigations where no existing tools are available, allowing forensic analysts to create customised solutions on demand. Integrating AI agents with these models could streamline evidence handling by allowing investigators to perform complex queries more intuitively, such as retrieving all messages from a specific date without the need to craft regular expressions.

In essence, while LLMs offer exciting prospects for the future of digital forensics, a balanced approach that integrates their strengths with human oversight is essential for harnessing their full potential. Inevitably, LLM-facilitated DF processes themselves will become the focus of future investigation.

\bibliographystyle{elsarticle-num-names}
\bibliography{refs}

\end{document}